\documentclass[AMA,STIX1COL]{WileyNJD-v2}

\articletype{Article Type}%

\received{}
\revised{}
\accepted{}

\begin{document}
	
	\title{Robust Estimation and Variable Selection for the Accelerated Failure Time Model}
	
	\author[1]{Yi Li*}
	
	\author[2]{Muxuan Liang}
	
	\author[3]{Lu Mao}
	
	\author[4]{Sijian Wang}

	\authormark{Li, Liang, Mao and Wang}

	\address[1]{\orgdiv{Department of Statistics}, \orgname{University of Wisconsin-Madison}, \orgaddress{\state{Wisconsin}, \country{USA}}}
	
	\address[2]{\orgdiv{Public Health Sciences Division}, \orgname{Fred Hutchinson Cancer Research Center}, \orgaddress{\state{Washington}, \country{USA}}}
	
	\address[3]{\orgdiv{Department of Biostatistics and Medical Informatics}, \orgname{University of Wisconsin-Madison}, \orgaddress{\state{Wisconsin}, \country{USA}}}
	
	\address[4]{\orgdiv{Department of Statistics}, \orgname{Rutgers University}, \orgaddress{\state{New Jersey}, \country{USA}}}

	\corres{*Yi Li, \email{li539@wisc.edu}}
	
	\presentaddress{Yi Li, 1300 University Ave, Madison, WI 53706}
	
	\abstract[Summary]{This paper considers robust modeling of the survival time for cancer patients. Accurate prediction can be helpful for developing therapeutic and care strategies. We propose a unified Expectation-Maximization approach combined with the $L_1$-norm penalty to perform variable selection and obtain parameter estimation simultaneously for the accelerated failure time model with right-censored survival data. Our approach can be used with general loss functions, and reduces to the well-known Buckley-James method when the squared-error loss is used without regularization. To mitigate the effects of outliers and heavy-tailed noise in the real application, we advocate the use of robust loss functions under our proposed framework. Simulation studies are conducted to evaluate the performance of the proposed approach with different loss functions, and an application to an ovarian carcinoma study is provided. Meanwhile, we extend our approach by incorporating the group structure of covariates.}
	
	\keywords{Cancer study; Censored data; Kaplan-Meier estimator; LASSO; Predictive robust regression.}

	\maketitle

\section{Introduction}\label{sec:intro}
The prediction of patient-specific survival time is an important problem in
cancer research. Accurate prediction using patient demographic, clinical, and genetic information
allows for targeted medication and care strategies.
Although the Cox proportional hazards regression \cite{cox} has been most popular in modeling the relationship
between a failure time and a set of predictors, it focuses on a summary quantity, i.e., hazard function,
of the outcome rather than the survival time itself. As a result, 
it is ill-suited for patient-level prediction. On the other hand, the accelerate failure time (AFT) model  \cite{aft1,aft2}
makes a much better tool in this regard as it models (the logarithm of) the survival time directly and explicitly 
in terms of a linear combination of covariates and an error term.
Prediction is thus made easy by plugging in the covariate values in question under a fitted
AFT model.
A common approach to fitting the AFT model with censored data is the Buckley-James method \cite{BJ},
which estimates the regression parameters by iteratively updating the error distribution using the Kaplan-Meier estimator \cite{KM} and the regression coefficients using a weighted least-squares.
%The common approaches to fitting the AFT model with censored data include the Stute's weighted least squares method \cite{stute1993,stute1996} and Buckley-James method \cite{BJ}. The Stute's weighted least squares method uses Kaplan-Meier weights \cite{KM} to account for censoring. The Buckley-James method imputes the censored observations by their conditional expectations given corresponding censoring times and covariates.

The traditional AFT model, however, often proves inadequate in this era of Big Data. 
With the advent of 
high-throughput technologies in molecular biology, it is possible to generate large volumes of gene
sequencing/expression data per patient that are potentially prognostic of survival.
Due to the curse of dimensionality, a model that exhausts all potential predictors often leads to over-fitting and thus poor performance.
Therefore, to leverage the abundance of (genetic) information for prediction, 
it is imperative to combine the traditional statistical model with a variable selection
procedure so that only those that are most and truly predictive of the outcome are included as covariates.
The selection process is typically accomplished via penalized regression, whereby a subset of regression coefficients
are shrunk to exact zeros, removing the corresponding covariates from the pool of predictors.
Popular choices for the penalty function include the least absolute shrinkage and selection operator (LASSO) \cite{lasso}, i.e., the $L_1$-norm, and its generalizations such as the adaptive LASSO \cite{alasso}, the smoothly clipped absolute deviation (SCAD) \cite{SCAD}, and the elastic net \cite{enet}, among others. 

Due to its unique importance in the prediction of survival times,
variable selection in the AFT model is of great interest to statisticians
and has been studied by many in the literature. 
For example, Huang et al. \cite{HSH2006}
and Huang and Ma \cite{HS2010} adopted different variants of the LASSO penalty and used Stute's inverse weighting \cite{stute1993,stute1996} on the uncensored observations to adjust for the censoring bias.
Wang et al. \cite{wang2008} used the elastic net penalty combined with Buckley-James approach for estimation and variable selection.
All the aforementioned methods are based on objective functions in the form of penalized sum of squared difference between observed and predicted outcomes. However, the squared-error loss function is known to be sensitive to outliers or heavy-tailed noise. 
To address this limitation, a number of robust regression methods have been proposed as an alternative to the least squares.
These include the least absolute deviation (LAD) regression \cite{lad}, regression with Huber's criterion \cite{huber}, and regression with Tukey's biweight criterion \cite{tukey}. For fully observed data, regularized versions of these robust regression methods have been widely studied. Wang et al. \cite{LAD-lasso} combined the LAD criterion and the LASSO-type penalty together to produce the LAD-LASSO method. Lambert-Lacroix and Zwald \cite{H-lasso} proposed to combine Huber's criterion and the adaptive LASSO. Chang et al. \cite{T-lasso} studied the Tukey-LASSO method, which combined Tukey's biweight criterion and the adaptive LASSO penalty. For censored data, robust loss functions such as the quantile loss have been applied to quantile regression models \cite{Quantile34,Quantile35,Quantile37}. However, to our knowledge, variable selection with a general loss function for the AFT model has not been considered in the literature.

In this paper, we propose a unified Expectation-Maximization (EM) approach combined with the $L_1$-norm penalty for variable selection and parameter estimation simultaneously for the AFT model with right-censored survival data. Our approach accommodates general loss functions, including those that are robust with regard to outliers and heavy-tailed errors. The Buckley-James method is a special case of our proposed approach when the squared-error loss is used without regularization.

The rest of the paper is organized as follows. In Section \ref{sec:our_method}, we first re-formulate the
Buckley-James method as an EM-type algorithm which motivates our proposed approach. Then, we present our unified regularized EM approach and its applications with robust loss functions. In Section \ref{sec:sim}, we conduct simulation studies to evaluate the performance of our approach with different loss functions. We apply the proposed approach to analyze data from an ovarian cancer study in Section \ref{sec:real}. In Section \ref{sec:aft_sgl}, we extend our approach by incorporating the group structure of covariates and use simulations to demonstrate the performance. Concluding remarks are given in Section \ref{sec:conclude}. 

\section{Theory and methods}\label{sec:our_method}

\subsection{A re-formulation of the Buckley-James as an EM-type algorithm}\label{sec:BJ}
We first cast the familiar Buckley-James estimator for the AFT
model in the framework of EM algorithm.
This re-formulation will offer us insights in generalizing the method to more robust approaches.

Let $T_i$ denote the logarithm of the survival time and $X_i$ denote the covariate vector for the $i$th subject. 
The accelerated failure time (AFT) model assumes that $T_i$ is linearly related to $X_i$, that is,
\begin{align}\label{eq:aft}
T_i= \alpha + X_i^T \beta + \epsilon_i, \ i=1,\ldots,n,
\end{align}
where $\alpha$ is the intercept, $\beta$ is the regression coefficient vector, and $\epsilon_i$ is a mean-zero error term independent with $X_i$.
In practice, the survival time is subject to
right censoring.
Let $C_i$ denote the logarithm of the censoring time and assume that $C_i\perp T_i\mid X_i$.
The observed data thus consist of $(Y_i,\delta_i,X_i)$, where $Y_i=\text{min}(T_i,C_i)$ and $\delta_i=1\{T_i \leq C_i\}$.
Had the $T_i$ been fully observed, model \eqref{eq:aft} would reduce to an ordinary linear
regression model and the ordinary least squares would apply.
In the presence of censoring, Buckley and James \cite{BJ} proposed
to supplement the least squares with an ``imputation step'' on the censored outcomes
in an iterative way.
The specific algorithm proceeds as follows. \\

\textit{Algorithm A:} The Buckley-James method
\begin{enumerate}
	\item[(A1)] Initialize $\alpha^{(0)}$ and $\beta^{(0)}$. Denote $X_i^{*}=X_i-\bar{X}$, where $\bar{X}=\sum_{i=1}^{n}X_i$, and $X^*=\left(X_1^*,\ldots,X_n^*\right)^T$.
	
	\item[(A2)] In the $m$th iteration, define the "imputed" survival time $Y_i^{(m)}$ by
	\begin{align}\label{eq:yi}
	Y_i^{(m)}=\delta_i Y_i+(1-\delta_i)E\left(T_i|\alpha^{(m-1)},\beta^{(m-1)},T_i>Y_i, Y_i\right), 
	\end{align}
	where the conditional expectation is computed by integration under a Kaplan-Meier estimator
	of the error distribution. Write $Y^{(m)}=(Y_1^{(m)},\ldots,Y_n^{(m)})^T.$
	
	\item[(A3)] Update $$\beta^{(m)} = \left({X^{*}}^T X^*\right)^{-1} {X^{*}}^T Y^{(m)}$$ 
	and 
	$$\alpha^{(m)}=\frac{1}{n}\sum_{i=1}^n Y_i^{(m)} - \bar{X}^T\beta^{(m)}.$$
	
	\item[(A4)] Repeat Steps (A2) and (A3) until convergence.
\end{enumerate}

The Buckley-James method can be viewed as an EM-type algorithm \cite{EM}. 
Consider minimizing the objective function
\begin{align}\label{eq:obj_fun}
l_n(\theta)&= \sum_{i=1}^{n}(T_i-\alpha-X_i^T\beta)^2\notag\\
&=  \sum_{i\in {\cal C}} (T_i-\alpha-X_i^T\beta)^2 +\sum_{i\in {\cal D}}(T_i-\alpha-X_i^T\beta)^2,
\end{align}
where $\theta=(\alpha,\beta^T)^T$, $\cal C$ is the index set for
the censored observations, and $\cal D$ is that for the uncensored ones. 
Clearly, the second term on the far right hand side of \eqref{eq:obj_fun} is computable while the first 
is not due to incomplete observation of the $T_i$.
However, one can use an ``E-step'' to impute the missing values therein.
At the $m$th iteration, the conditional expectation of $l_n(\theta)$
given the observed data and the parameters from the last iteration is
$$
Q(\theta|\theta^{(m-1)})=E\{l_n(\theta)\mid\theta^{(m-1)},\mbox{observed data}\}=\sum_{i\in {\cal C}}E_{T_i}\left\{(T_i-\alpha-X_i^T\beta)^2|\theta^{(m-1)}, T_i>Y_i, Y_i\right\}+\sum_{i\in {\cal D}}(Y_i-\alpha-X_i^T\beta)^2.
$$
Then, at the ``M-step'', we obtain $\theta^{(m)}$ by
$$\theta^{(m)} = \operatorname*{argmin}_\theta Q(\theta|\theta^{(m-1)}).$$
To find an explicit expression for $\theta^{(m)}$,
set $\partial Q(\theta|\theta^{(m-1)})/\partial\theta=0$ to obtain
\begin{align}\label{eq:m_step}
\begin{array}{ll}
\sum_{i\in {\cal C}}\left\{E(T_i|\theta^{(m-1)}, T_i>Y_i, Y_i)-\alpha - X_i^{T}\beta\right\} + \sum_{i\in {\cal D}}(Y_i-\alpha - X_i^{T}\beta) = 0, \\
\sum_{i\in {\cal C}}\left\{E(T_i|\theta^{(m-1)}, T_i>Y_i, Y_i)-\alpha - X_i^{T}\beta\right\} X_i + \sum_{i\in {\cal D}}(Y_i-\alpha - X_i^{T}\beta) X_i = 0.
\end{array}  
\end{align}
With $Y_i^{(m)}$ defined in \eqref{eq:yi} in Step (A2) , the equations in \eqref{eq:m_step}
reduce to
$$
\Bigg\{ \begin{array}{ll}
\sum_{i=1}^{n} (Y_i^{(m)}-\alpha - X_i^{T}\beta) =0, \\
\sum_{i=1}^{n} (Y_i^{(m)}-\alpha - X_i^{T}\beta) X_i=0,
\end{array}  
$$
leading to Step (A3) of the Buckley-James method.
Thus, we have re-formulated the Buckley-James as an EM-type algorithm.

When the errors are normally distributed,  the squared-error loss function $l_n(\theta)$ defined in \eqref{eq:obj_fun} is proportional to the negative log-likelihood. In such cases, the Buckley-James corresponds to a maximum likelihood estimator obtained from a true EM algorithm
and is thus likely to have desirable properties.
However, when the errors are non-Gaussian, especially when they are heavy-tailed or contain outliers, the squared-error loss need not be appropriate.  A similar EM-type algorithm but with alternative choices of $l_n(\theta)$ may yield a more robust procedure.

%Our approach is flexible with not only loss functions but also penalty terms. We could use the sparse group lasso (SGL) penalty \cite{SGL} to perform variable selection at both the group level and the within-group level simultaneously. More details are described in the Supplementary Material.

\subsection{A unified framework for regularized EM estimation}\label{sec:our}
We generalize the EM-type approach of the Buckley-James method described in Section \ref{sec:BJ}
in two respects. First, we replace the squared-error loss with a general loss function. Second, we add to the objective function a regularization, or penalty, term for the regression coefficients so as to achieve variable selection.

Let $L(T_i,\theta)$ be a general loss function of interest, 
where for simplicity we have suppressed its dependence on $X_i$.
To incorporate the $L_1$ regularization in the framework of the EM algorithm, consider minimizing the following objective function with respect to $\theta$:
\begin{align}\label{eq:obj_general}
l_n(\theta;\lambda)=&\sum_{i=1}^{n}L(T_i,\theta)+\lambda \sum_{j=1}^{p}|\beta_j| \notag\\
=&  \sum_{i\in {\cal C}} L(T_i,\theta)+\sum_{i\in {\cal D}}L(T_i,\theta)+\lambda \sum_{j=1}^{p}|\beta_j|,
\end{align}
where $\lambda\geq 0$ is the tuning parameter. 
The difference of \eqref{eq:obj_general} with \eqref{eq:obj_fun} lies in
the replacement of the squared-error loss function by a general loss function
$L$ and the addition of an $L_1$ penalty term for $\beta$.
The key idea in the solution to the minimization problem, however,
remains the same: to apply the EM algorithm treating the censored observations as missing data. 

Given $\theta^{(m-1)}$ from the $(m-1)$th iteration, the Q-function in the ``E-step'' of the $m$th iteration is
\begin{align}\label{eq:Q_fun}
Q(\theta|\theta^{(m-1)})&=E\{l_n(\theta;\lambda)\mid\theta^{(m-1)},\mbox{observed data}\}\notag\\
&=\sum_{i\in {\cal C}}E_{T_i}\left\{L(T_i,\theta)|\theta^{(m-1)}, T_i>Y_i, Y_i\right\}+\sum_{i\in {\cal D}}L(Y_i,\theta) + \lambda \sum_{j=1}^{p}|\beta_j|,
\end{align}
More specifically, the conditional expectation $E_{T_i}\left\{L(T_i,\theta)|\theta^{(m-1)}, T_i>Y_i, Y_i\right\}$ in \eqref{eq:Q_fun} can be calculated as follows. Let $\xi_i^{(m-1)} = T_i - \alpha^{(m-1)} - X_i^T\beta^{(m-1)}$. Then we have
\begin{align}\label{eq:ce_cal}
&E_{T_i}\left\{L(T_i,\theta)|\theta^{(m-1)}, T_i>Y_i, Y_i\right\}\notag\\
&=E_{\xi_i^{(m-1)}}\left\{L(\xi_i^{(m-1)}+\alpha^{(m-1)}+X_i^T\beta^{(m-1)},\theta)|\theta^{(m-1)}, \xi_i^{(m-1)}>Y_i- \alpha^{(m-1)} - X_i^T\beta^{(m-1)}, Y_i\right\}\notag\\
&=\frac{\int_{Y_i- \alpha^{(m-1)} - X_i^T\beta^{(m-1)}}^{\infty}L(t+\alpha^{(m-1)}+X_i^T\beta^{(m-1)},\theta)dF^{(m-1)}(t)}{1-F^{(m-1)}(Y_i- \alpha^{(m-1)} - X_i^T\beta^{(m-1)})},
\end{align}
where $F^{(m-1)}(t)$ is the cumulative distribution function of  $\xi_i^{(m-1)}$.

We then estimate $F^{(m-1)}(t)$ by the Kaplan-Meier estimator based on the $\xi_i^{(m-1)}$. Denote $e_i^{(m-1)} = Y_i - \alpha^{(m-1)} - X_i^T\beta^{(m-1)}, \ i=1,\ldots,n$. Without loss of generality, we assume that $e_1^{(m-1)} \leq \ldots \leq e_n^{(m-1)}$. Otherwise, we could sort $e_i^{(m-1)}$'s in ascending order and re-arrange the $(Y_i,\delta_i,X_i)$ accordingly. In case of ties between censored and uncensored observations, put uncensored observations before censored ones. As suggested by Buckley and James \cite{BJ}, one always picks an uncensored observation for the largest residual. That is, $e_n^{(m-1)}$ is always uncensored. The resulting estimator of $F^{(m-1)}(t)$ can be written as  
\begin{align*}
\widehat{F}^{(m-1)}(t)=&\sum_{i:e_i^{(m-1)} \leq t} m_i^{(m-1)}\\
=&\sum_{i=1}^{n}m_i^{(m-1)}1\{e_i^{(m-1)} \leq t\},
\end{align*}
where $m_i^{(m-1)}$'s are called the Kaplan-Meier weights and can be expressed as
$$
m_1^{(m-1)}=\frac{\delta_1}{n}, \ \  m_i^{(m-1)}=\frac{\delta_i}{n-i+1}\prod_{j=1}^{i-1} \left(\frac{n-j}{n-j+1}\right)^{\delta_j}, \ i=2,\ldots,n.
$$
As a result, the right hand side of \eqref{eq:ce_cal} can be estimated by
\begin{align}\label{eq:ce_result}
&\frac{\int_{Y_i- \alpha^{(m-1)} - X_i^T\beta^{(m-1)}}^{\infty}L(t+\alpha^{(m-1)}+X_i^T\beta^{(m-1)},\theta)d\widehat{F}^{(m-1)}(t)}{1-\widehat{F}^{(m-1)}(Y_i- \alpha^{(m-1)} - X_i^T\beta^{(m-1)})}\notag\\ 
&= \frac{\sum\limits_{u>i} m_u^{(m-1)} L(e_u^{(m-1)}+\alpha^{(m-1)}+X_i^T\beta^{(m-1)},\theta)}{\sum\limits_{u>i} m_u^{(m-1)}}\notag\\
&=\sum_{u>i}w_{iu}^{(m-1)}L(e_u^{(m-1)}+\alpha^{(m-1)}+X_i^T\beta^{(m-1)},\theta),
\end{align}
where $w_{iu}^{(m-1)}=m_u^{(m-1)}/\sum_{u>i} m_u^{(m-1)}$ for $u>i$. Therefore, we can estimate Q-function in \eqref{eq:Q_fun} by 	
\begin{align}\label{eq:Q_final}
\widehat{Q}(\theta|\theta^{(m-1)})=&
\sum_{i\in C}\sum_{u>i}w_{iu}^{(m-1)}L(e_u^{(m-1)}+\alpha^{(m-1)}+X_i^T\beta^{(m-1)},\theta)\notag\\
&+\sum_{i\in D} L(Y_i,\theta)+\lambda \sum_{j=1}^{p}|\beta_j|. 
\end{align}

Now, for the ``M-step'', $\theta^{(m)}$ is updated by computing
$$\theta^{(m)} = \operatorname*{argmin}_\theta \widehat{Q}(\theta|\theta^{(m-1)})$$ 
 using suitable numerical algorithms.

Overall, the proposed regularized EM approach for the AFT model 
can be summarized in the following algorithm. \\

\textit{Algorithm B:} The regularized EM approach for the AFT model
\begin{enumerate}
	%\item Sort $Y_i$'s in ascending order and arrange $(\delta_i,X_i)$'s according the ordered $Y_i$'s. Let $\delta_n  = 1$.
	
	\item[(B1)] Initialize $\theta^{(0)}$.
	
	\item[(B2)] In the $m$th iteration, calculate $\widehat{Q}(\theta|\theta^{(m-1)})$ by
	\begin{align*}
	\widehat{Q}(\theta|\theta^{(m-1)})=&
	\sum_{i\in C}\sum_{u>i}w_{iu}^{(m-1)}L(e_u^{(m-1)}+\alpha^{(m-1)}+X_i^T\beta^{(m-1)},\theta)\\
	&+\sum_{i\in D} L(Y_i,\theta)+\lambda \sum_{j=1}^{p}|\beta_j|. 
	\end{align*}
	
	\item[(B3)] Update $\theta^{(m)}$ by minimizing $\widehat{Q}(\theta|\theta^{(m-1)})$.
	
	\item[(B4)] Repeat Steps (B2) and (B3) until the convergence criterion is met. 
\end{enumerate}

We use $\|\theta^{(m)}-\theta^{(k)}\|_{\infty}<10^{-5}$ for some $k$ as the convergence criterion. The rationale for such convergence criterion is to circumvent possible oscillations as iteration proceeds (for more details, see Wang et al. \cite{wang2008} and references therein). Following the strategy in Wang et al. \cite{wang2008}, we take the estimate from the last iteration as the final solution.

\subsection{Applications with robust loss functions}\label{sec:robust}

To illustrate our general approach with specific loss functions $L$, we consider three robust loss functions: the absolute loss, the Huber loss and Tukey's biweight loss (the Tukey loss in short).

%We first calculate and sort $e_i^{(m-1)}$'s in ascending order and re-arrange the $(Y_i,\delta_i,X_i)$ accordingly.
For notational simplicity, we can re-formulate $\widehat{Q}(\theta|\theta^{(m-1)})$
in terms of an augmented dataset. For $i \in \cal C$, let $c_i$ be the number of uncensored observations with indices larger than $i$ and $(u_{i1},\ldots,u_{ic_i})$ denote these $c_i$ indices. Define $c_i$ new data points as
\begin{align*}
Y_{i(v)}^{new}=e_{u_{iv}}^{(m-1)}+\alpha^{(m-1)}+X_i^T\beta^{(m-1)}, \ \ X_{i(v)}^{new}=X_i, \ \ w_{i(v)}^{new}=w_{i{u_{iv}}}^{(m-1)}, \ \ v=1,\ldots,c_i.
\end{align*}
For $ i \in {\cal D}$, set $c_i=1$ and define
\begin{align*}
Y_{i(1)}^{new}=Y_i, \ \ X_{i(1)}^{new}=X_i, \ \ w_{i(1)}^{new}=1.
\end{align*}
Then
\begin{align}\label{eq:Q_est}
\widehat{Q}(\theta|\theta^{(m-1)})=\sum_{i=1}^n\sum_{v=1}^{c_i} w_{i(v)}^{new}L(Y_{i(v)}^{new},X_{i(v)}^{new},\theta) +\lambda \sum_{j=1}^{p}|\beta_j|.
\end{align}
Denote $K = \sum_{i=1}^n c_i $, we can rewrite the double summation in \eqref{eq:Q_est} as a single summation
\begin{align}\label{eq:aft_robust}
\widehat{Q}(\theta|\theta^{(m-1)})=\sum_{k=1}^{K} w_{k}^{new}L(Y_{k}^{new},X_{k}^{new},\theta) +\lambda \sum_{j=1}^{p}|\beta_j|,
\end{align}
where there exists $i$ and $v$ satisfying $\sum_{i'=1}^{i-1}c_{i'}+v=k$ such that $(Y_{k}^{new}, X_{k}^{new}, w_{k}^{new}) = (Y_{i(v)}^{new}, X_{i(v)}^{new}, w_{i(v)}^{new})$.

The absolute loss function is given by $L(T_i,\theta)=|T_i-\alpha-X_i^T\beta|$. This leads to 
\begin{align} \label{eq:aft_ab}
\widehat{Q}(\theta|\theta^{(m-1)})=\sum_{k=1}^{K} w_{k}^{new}| Y_{k}^{new} -\alpha - X_{k}^{new^T} \beta| +\lambda \sum_{j=1}^{p}|\beta_j| .
\end{align}
To minimize $\widehat{Q}(\theta|\theta^{(m-1)})$,  we first re-formulate the data set $(Y_i^{\#},X_i^{\#})$ as 
\begin{align*}
(Y_i^{\#},X_i^{\#}) = \Bigg\{ \begin{array}{ll}
( Y_{i}^{new},  X_{i}^{new}) &    \text{for} \ i=1,\ldots,K , \\
( 0,  \lambda e_{i-K}) &   \text{for} \ i= K+1,\ldots, K+p , 
\end{array}   
\end{align*}
where $e_j$ is the $p$-dimensional vector with the $j$th component being 1 and all others being 0. The minimization problem can be written in terms of artificial variables $v_i$ as
\begin{align*}
\min_{\alpha,\beta,v} \left(\sum_{i=1}^{K} w_{i}^{new} v_i + \sum_{i=K+1}^{K+p} v_i \right)  \ \  \text{subject to} \ \  & -v_i \leq Y_i^{\#}-\alpha-X_i^{\#^T}\beta \leq v_i \ \ \text{ for }i=1,\ldots,K, \\
& -v_i \leq Y_i^{\#} - X_i^{\#^T}\beta \leq v_i \ \ \text{ for }i=K+1,\ldots,K+p.
\end{align*}
Then the simplex algorithm \cite{lad_a} can be used to solve the above linear programming.

The Huber loss function $L^{\text{Huber(h)}}(x)$ proposed by Huber is an elegant compromise between the squared-error loss function and the absolute loss function. It is defined as 
\begin{align}\label{eq:huber}
L^{\text{Huber(h)}}(x)= \Bigg\{ \begin{array}{ll}
x^2 &    \text{if} \ |x| \leq h , \\
h(2|x|-h)&   \text{if} \ |x| > h , 
\end{array}  
\end{align}
where $h$ is a parameter which controls the amount of robustness. This function is quadratic when $|x| \leq h$ and linear as $|x| > h$. For the Huber loss, $\widehat{Q}(\theta|\theta^{(m-1)})$ can be written as
%The Huber loss converges to the absolute loss as $k$ tends to 0, and it converges to the squared loss as $k$ tends to $+\infty$. For the Huber loss, $Q(\theta|\theta^{(m-1)})$ can be written as 
\begin{align}\label{eq:aft_huber}
\widehat{Q}(\theta|\theta^{(m-1)})=\sum_{k=1}^{K} w_{k}^{new} L^{\text{Huber(h)}}\left(\frac{Y_{k}^{new} - \alpha - X_{k}^{new^T}\beta}{d}\right) +  \lambda \sum_{j=1}^{p} \frac{|\beta_j|}{d},
\end{align}
where $d$ is a scale parameter and can be estimated with the regression coefficients simultaneously. 

\begin{remark}
Huber \cite{huber1973,huber2011robust} suggested scaling the residuals in order to make the resulting estimator scale-invariant. Following the suggestion, we scale the residuals and the penalty term in \eqref{eq:aft_huber} to obtain a scale-invariant estimator. One of the advantages of scale-invariance is that the results do not depend on the measurement units. For the absolute loss, scaling is not necessary since the objective function is homogeneous with respect to the residuals and the $L_1$-norm of the regression coefficients. 
\end{remark}
 
%In the above formulation, the residuals and the penalty term are scaled so that the Huber estimator is scale-invariant (see Huber \cite{huber1973,huber2011robust}). One of the advantages of scale-invariance is that the results do not depend on the measurement units. For the absolute loss, scaling is not necessary since the objective function is homogeneous with respect to the residuals and the $L_1$-norm of the regression coefficients. 

%For the squared-error loss, the conventional does not consider the scaling, but the tuning parameter of the penalty term can be adjusted to yield an equivalent scale-invariant solution. By the same reasoning, it is possible to adjust the tuning parameter of the penalty term in forming the Huber estimator to achieve a scale-invariant solution, but here we want to provide an explicit formulation to emphasize that our estimator is scale-invariant. We will do the same for the Tukey estimator.

The scale parameter $d$ can be estimated by the normalized weighted median absolute deviation of the residuals. For the robustness parameter $h$, Huber\cite{huber2011robust} recommended using $h=1.345$,
which yields a 95\% asymptotic relative efficiency compared with the least squares solution 
under normally distributed errors.
%More details about the selection of $h$ and the estimation of $d$ are provided in Appendix A. 

Another well-known robust loss function is the Tukey loss $L^{\text{Tukey(t)}}(x)$ defined by
\begin{align}\label{eq:tukey}
L^{\text{Tukey(t)}}(x)= \Bigg\{ \begin{array}{ll}
\frac{t^2}{6}\left[1-\left\{1-(\frac{x}{t})^2\right\}^3\right] &    \text{if} \ |x| \leq t , \\
\frac{t^2}{6}    &       \text{if} \ |x| > t,
\end{array} 
\end{align}
where $t$ is a robustness parameter. Similarly, for the Tukey loss we have 
\begin{align}\label{eq:aft_tukey}
\widehat{Q}(\theta|\theta^{(m-1)})=\sum_{k=1}^{K} w_{k}^{new} L^{\text{Tukey(t)}}\left(\frac{Y_{k}^{new} - \alpha - X_{k}^{new^T}\beta}{d}\right) + \lambda \sum_{j=1}^{p} \frac{|\beta_j|}{d},
\end{align}
where $d$ is still a scale parameter. The parameter $t$ in \eqref{eq:tukey} is recommended to take the value 4.685 for the same reason as the suggested value $h$ in \eqref{eq:huber}. Similar to the Huber case, we scale the residuals and the penalty term in \eqref{eq:aft_tukey} as well.

%We use an iterative coordinate descent algorithm to solve the penalized weighted Huber and Tukey regressions. The algorithm shares the same idea as that proposed by Mkhadri et al. \cite{cdaSQR} for computing the penalized smooth quantile regression (cdaSQR) where the Majorization-Minimization trick was adopted. Specifically, a majoring quadratic function of the objective function is first constructed and then a coordinate descent algorithm is used to optimize this quadratic approximation. Mkhadri et al. also established the convergence property of the cdaSQR approach with the elastic net penalty. 
%

We use the cdaSQR algorithm proposed by Mkhadri et al. \cite{cdaSQR} to solve the penalized weighted Huber and Tukey regressions. The cdaSQR algorithm is an iterative coordinate descent algorithm for computing the penalized smooth quantile regression where the Majorization-Minimization trick was adopted. Specifically, a majoring quadratic function of the objective function is first constructed and then a coordinate descent algorithm is used to optimize this quadratic approximation. This process is repeated until convergence. Mkhadri et al. \cite{cdaSQR} also established the convergence property of the cdaSQR approach with the elastic net penalty. 

The iterative coordinate descent algorithm proceeds as follows. Let $\widetilde{\alpha}$ and $\widetilde{\beta}$ denote the current values of the parameters. Then, the objective function to be minimized for each component $\beta_j, \ j=1,\ldots,p$, can be written as
\begin{align}\label{eq:obj_ht}
\sum_{k=1}^{K} w_{k}^{new} \phi \left\{ \frac{r_k + X_{kj}^{new}(\widetilde{\beta_j} - \beta_j)}{d}\right\} + \lambda \frac{|\beta_j|}{d},
\end{align}
where $r_k = Y_{k}^{new} - \widetilde{\alpha} - X_{k}^{new^T} \widetilde{\beta}$ and $\phi$ is the Huber loss $L^{\text{Huber(h)}}$ or the Tukey loss $L^{\text{Tukey(t)}}$.

A majoring quadratic function of \eqref{eq:obj_ht} is given by
\begin{align}\label{eq:obj_ht_approx}
\sum_{k=1}^{K} w_{k}^{new} \phi\left(\frac{r_k}{d}\right) + 
\sum_{k=1}^{K} w_{k}^{new} \phi'\left(\frac{r_k}{d}\right) \frac{X_{kj}^{new}(\widetilde{\beta_j} - \beta_j)}{d} + 
\sum_{k=1}^{K} w_{k}^{new} \left\{\frac{X_{kj}^{new}(\widetilde{\beta_j} - \beta_j)}{d} \right\}^2   + 
\lambda \frac{|\beta_j|}{d},
\end{align}
where $\phi'$ is the derivative of $\phi$.

Let $S(z,\gamma) = \text{sgn}(z)(|z|-\gamma)_+$ be the soft-thresholding operator. Then the coordinate-wise update to minimize \eqref{eq:obj_ht_approx} has the form
\begin{align*}
\beta_j \leftarrow \frac{S\left\{ 2 \sum_{k=1}^{K} w_{k}^{new} \frac{X_{kj}^{new^2}}{\hat{d}^2}\widetilde{\beta_j} + 
	\sum_{k=1}^{K} w_{k}^{new} \phi'(\frac{r_k}{\hat{d}}) \frac{X_{kj}^{new}}{\hat{d}}, \frac{\lambda}{\hat{d}}\right\} }
{2\sum_{k=1}^{K} w_{k}^{new} \frac{X_{kj}^{new^2}}{\hat{d}^2}},
\end{align*}
where $\hat{d} = \frac{\text{WMedian}\left(|r_k-\text{WMedian}(r_k,w_{k}^{new})|, w_{k}^{new}\right)}{0.6745}$ and 
$\text{WMedian}(r_k,w_{k}^{new}):=\operatorname*{argmin}_u \sum_{k = 1}^{K}w_{k}^{new} |r_k-u|$ is the weighted median.
Similarly, we can update the intercept term by
\begin{align*}
\alpha \leftarrow \widetilde{\alpha} + \frac{\sum_{k=1}^{K} w_{k}^{new} \phi'(\frac{r_k}{\hat{d}})}{2\sum_{k=1}^{K} w_{k}^{new}}\hat{d}.
\end{align*}
As a result, the iterative coordinate descent algorithm to minimize the objective functions \eqref{eq:aft_huber} and  \eqref{eq:aft_tukey} can be summarized as follows.\\

\textit{Algorithm C:} The iterative coordinate descent algorithm for the penalized weighted Huber and Tukey regressions with the LASSO penlaty
\begin{enumerate}
	\item[(C1)] Initialize $\widetilde{\alpha}$ and $\widetilde{\beta}$.
	
	\item[(C2)] Update the regression coefficients: for $j=1,\ldots,p$
		\begin{enumerate}
			\item Calculate $r_k = Y_{k}^{new} - \widetilde{\alpha} - X_{k}^{new^T} \widetilde{\beta}$ \ for $k=1,\ldots,K$.
			\item Update $$\widehat{\beta_j} = \frac{S\left\{ 2\sum_{k=1}^{K} w_{k}^{new} \frac{X_{kj}^{new^2}}{\hat{d}^2}\widetilde{\beta_j} + 
				\sum_{k=1}^{K} w_{k}^{new} \phi'(\frac{r_k}{\hat{d}}) \frac{X_{kj}^{new}}{\hat{d}}, \frac{\lambda}{\hat{d}}\right\}}
			{ 2\sum_{k=1}^{K} w_{k}^{new} \frac{X_{kj}^{new^2}}{\hat{d}^2}},$$ where $\hat{d} = \frac{\text{WMedian}\left(|r_k-\text{WMedian}(r_k,w_{k}^{new})|, w_{k}^{new}\right)}{0.6745}$.
			\item Set $\widetilde{\beta_j} = \widehat{\beta_j}$.
		\end{enumerate}
	
	\item[(C3)] Update the intercept term:
		\begin{enumerate}
			\item Calculate $r_k = Y_{k}^{new} - \widetilde{\alpha} - X_{k}^{new^T} \widetilde{\beta}$ \ for $k=1,\ldots,K$.
			\item Update $$\widehat{\alpha} = \widetilde{\alpha} + \frac{\sum_{k=1}^{K} w_{k}^{new} \phi'(\frac{r_k}{\hat{d}})}{2\sum_{k=1}^{K} w_{k}^{new}}\hat{d},$$
			where $\hat{d} = \frac{\text{WMedian}\left(|r_k-\text{WMedian}(r_k,w_{k}^{new})|, w_{k}^{new}\right)}{0.6745}$.
			\item Set $\widetilde{\alpha} = \widehat{\alpha}$.
		\end{enumerate}
	
	\item[(C4)] Repeat Steps (C2) and (C3) until convergence.
	
\end{enumerate}

\section{Simulation studies}\label{sec:sim}
In this section we conduct simulation studies to evaluate the performance of our approach with four different loss functions: (1) the squared-error loss; (2) the absolute loss; (3) the Huber loss; (4) the Tukey loss. We use $h=1.345$ in the Huber loss and $t=4.685$ in the Tukey loss as discussed in Section \ref{sec:robust}. We use a simulation study to show that we could still obtain the desired asymptotic relative efficiencies of the Huber estimator and the Tukey estimator with respect to the least squares estimator for survival outcomes. The detailed results are shown in the online Supporting Information.

The data are generated from the following model:
$$T_i = \alpha + X_i^{T}\beta + \sigma \epsilon_i, \ \ i=1,\ldots,n,$$ 
where $T_i$ is the logarithm of the true failure time, and $X_i$ is a $p$-dimensional covariate vector generated from the standard multivariate normal distribution ($N_p(\mathbf{0},\mathbf{I}_p)$). For the error terms, we consider the following different distributions: 
\begin{enumerate}
	\item the standard normal distribution $N(0,1)$,
	\item the mixture of two normal distributions $0.9N(0,1) + 0.1N(0,15^2)$.
	%: with probability 0.9, $\epsilon_i \sim N(0,1)$ and with probability 0.1,  $\epsilon_i \sim N(0,15^2)$,
	%\item the $t$-distribution with 3 degrees of freedom ($t_3$).
\end{enumerate}
The mixture of normal distributions generates outliers and is heavy-tailed. For different settings, $\sigma$ is chosen such that the signal-to-noise ratio (SNR), which is defined as $\frac{\beta^T \beta}{\sigma^2 \text{Var}(\epsilon_i)}$, equals 5 or 1. The censoring times are generated from a uniform distribution that yields about $30\%$ censoring rate. The number of covariates $p$ is 40. We set $\alpha=1$ and $\beta = (3,1.5,2,2,1,2,1.5,1,\underbrace{0,...,0}_{32})$, so that there are eight important covariates.
%For each training set, we generate an independent validation data set with the same sample size to select the tuning parameter $\lambda$. We consider two different tuning procedures. 
For each scenario, we generate three independent data sets: a training set to fit the model, a validation set to select the tuning parameter, and a test set to evaluate the prediction performance. We consider similar sample size as in the empirical analysis. The sample sizes of these three parts are 200, 100 and 200, respectively.

We consider two different tuning procedures. For the first tuning procedure, we only use uncensored observations in the validation set. Since the true failure times are known for uncensored observations, we define the validation error for a given $\lambda$ as $$VE_1=\frac{1}{m}\sum_{i=1}^{m} L(T_i,\hat{\theta}),$$ where $m$ is the number of uncensored observations and $\hat{\theta}$ is the estimator of $\theta$. We consider both non-refitting and refitting methods when we calculate $\hat{\theta}$. For the non-refitting method, we use the estimator directly from minimizing the penalized objective function. For the refitting method, after we select the important covariates, we re-estimate the coefficients of the selected covariates with no penalty. 
%we use the covariates selected by our method to re-calculate the estimator with no penalty. 

For the second tuning procedure, we use all observations, both uncensored and censored. We still use $L(T_i,\hat{\theta})$ to calculate the validation error for uncensored observations. For censored ones, we use $E_{T_i}\left\{L(T_i,\hat{\theta})|\hat{\theta},T_i>Y_i, Y_i\right\}$ which can be estimated as in \eqref{eq:ce_cal} and \eqref{eq:ce_result} to approximate the validation error. As a result, the validation error is defined as
$$
VE_2 = \frac{1}{n}\sum_{i=1}^{n} \left[(1-\delta_i)E_{T_i}\left\{L(T_i,\hat{\theta})|\hat{\theta},T_i>Y_i, Y_i\right\}+\delta_i L(T_i, \hat{\theta}) \right].
$$
Again, we consider both non-refitting and refitting methods when we calculate $\hat{\theta}$. We select the $\lambda$ that gives the lowest validation error.
%We fit the model by the proposed method with 100 different values of the tuning parameter $\lambda$. We choose the $\lambda$ that gives the lowest $VE$.
%We fit The tuning parameter which gives the lowest $VE$ is selected. We select the value of $\lambda$ that minimizes the validation error ($VE$).

To evaluate the performance with different loss functions, we consider the following four criteria:
\begin{enumerate}
	\item Sensitivity of selection (SEN): $\text{SEN} = \frac{\text{number of selected important covariates}}{\text{number of true important covariates}}$,
	\item Specificity of selection (SPE): 
	$\text{SPE} = \frac{\text{number of removed unimportant covariates}}{\text{number of true unimportant covariates}}$,
	\item Squared estimation error (SEE): SEE = $\|\hat{\beta} - \beta\|_2^2 + (\hat{\alpha} - \alpha)^2$,
	\item Prediction error (PE): PE = $n^{-1}\sum\limits_{i=1}^{n} \left\{(1-\delta_i)E_{T_i}\left(|T_i-\hat{\alpha} - X_i^T\hat{\beta}||\hat{\alpha},\hat{\beta},T_i>Y_i, Y_i\right)+\delta_i |T_i-\hat{\alpha} - X_i^T\hat{\beta}| \right\}$.
\end{enumerate}
The prediction error is calculated on the test set. In order to make the results comparable across different loss functions, we consider the absolute loss function as the common criterion here.
We also include the oracle method for comparison. Under the "oracle", we assume the true important covariates are known. We consider the estimation procedure with these covariates only, 
which theoretically produces the most favorable results one can hope for. Higher values of sensitivity and specificity and lower values of SEE and PE are desirable. We repeat each simulation setting 100 times, and calculate the means and standard errors of these four statistics.
Tables \ref{tab:normalsnr5} and \ref{tab:normalsnr1} summarize the results for the standard normal error with SNR=5 and 1. Tables \ref{tab:mixsnr5} and \ref{tab:mixsnr1} summarize the results for the normal mixture errors with SNR=5 and 1.

%the sensitivities of the three robust losses are higher whereas the specificity of four losses are similar.
We first look at the performance of variable selection. For the case of the standard normal errors, the squared-error loss is expected to perform well. Our simulations show that the squared-error loss performs better than the absolute loss. The Huber loss and the Tukey loss provide similar performance compared with the squared-error loss. For the case of errors with outliers, the three robust losses outperform the squared-error loss. We can observe that the three robust losses have higher sensitivity of selection than the squared-error loss whereas all four losses have similar specificity.
In terms of parameter estimation, the findings are similar. The Huber loss and the Tukey loss lead to similar SEE compared with the squared-error loss in the standard normal errors case, while the absolute loss yields the highest SEE. In the presence of outliers, the three robust losses produce much more accurate estimates. Due to accurate estimation, the three robust losses provide better prediction.

In addition, all loss functions improve their performance as the signal-to-noise ratio increases. The two tuning procedures have similar performance in the simulations. We will use tuning procedure 2 in the empirical analysis since it uses all observations. Furthermore, the performance of the refitting method is better than that of the non-refitting method when the signal is strong because all loss functions have better variable selection performance with strong signal.
Overall, the Huber loss and the Tukey loss perform similarly to the squared-error loss when the errors are normally distributed, but substantially better than the squared-error loss when outliers are present.

We also conducted simulation studies with errors following the $t$-distribution with 3 degrees of freedom ($t_3$). The results are summarized in the online Supporting Information. We can observe similar patterns as in Tables \ref{tab:normalsnr5}--\ref{tab:mixsnr1}.
%The observations are similar to what we made as.

\section{Real data analysis}\label{sec:real}
In this section, we apply the proposed approach to an ovarian carcinoma study which is part of the Cancer Genome Atlas (TCGA) project. Messenger RNA expression data from a total of 12,688 unique genes for 637 subjects are produced by microarray chips. In the study, death from any cause, often called overall survival, is the event of interest. Among the 637 subjects, 288 were censored. The original dataset consists of three parts, a training set of size 252, a validation set of size 134 and a test set of size 251. Due to the large amount of genes, screening is necessary to lower the noise level and to reduce the computational burden. We use the Somers' D \cite{C_index,D_index},
an asymmetric measure of ordinal association between two variables, to screen the genes. For each individual gene, the Somers' D between the gene expression and the failure time is calculated based on all 637 subjects. Following the recommendation in Fan and Lv \cite{SIS} and Mai and Zou \cite{non_SIS}, we screen the number of genes down to $d_n = [n/\log{n}] = [252/\log{(252)}] \approx 50$, which means that the genes with top 50 absolute Somers' D are retained in the analysis.

After the screening process, to make the results of variable selection and estimation more stable, we combine all three parts together
and then randomly divide the whole dataset into the above three parts with their corresponding sizes. We apply our approach with four different loss functions (the squared-error loss, the absolute loss, the Huber loss and the Tukey loss) to fit the training set. We use the second tuning criterion with the non-refitting method as discussed in Section \ref{sec:sim} since the non-refitting method will be more conservative in terms of variable selection. Then, the validation error is calculated on the validation set for each tuning parameter, and we select the tuning parameter that gives the lowest validation error. After that, we record the selected model and calculate the prediction error based on the test set. This whole procedure is repeated for 100 times.

The means and standard errors of PE for different loss functions are shown in Table \ref{tab:emp_predict_error}. It can be seen that the three robust losses have smaller PE than the squared-error loss.

\begin{figure}[!h]
	\centering
	\includegraphics[width=0.6\linewidth]{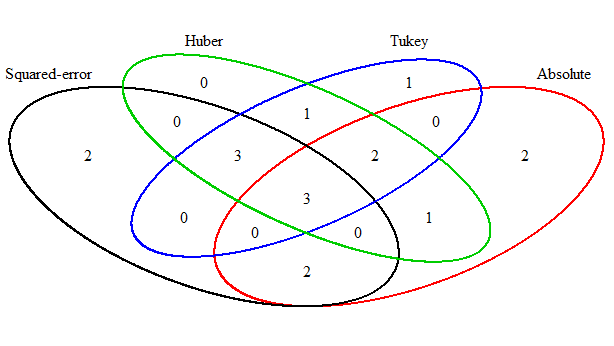}
	\captionof{figure}{The Venn diagram of the top 10 most frequently selected genes by different loss functions. Each number indicates the number of genes in the corresponding group.}
	\label{Fig:venn_new}
\end{figure}

Table \ref{tab:emp_top10_genes} gives the top 10 most frequently selected genes by different loss functions with their selection frequencies. In each column, the genes are listed in descending order of their selection frequencies. Figure \ref{Fig:venn_new} displays the Venn diagram of these four gene lists. In summary, three genes are selected by all four loss functions, among which VSIG4 has been confirmed to be overexpressed in ovarian cancers compared with that in benign tumors \cite{gene1}. Two genes (AKT2 and CLIP3) are identified by the three robust losses but not identified by the squared-error loss. In particular, it has been documented in the literature that AKT2 contributes to increased ovarian cancer cell migration and invasion \cite{gene2}. In addition, the gene BRD4, which is selected by the absolute loss and the Huber loss, has been identified in the literature as a potential therapeutic target in ovarian cancer \cite{gene3}. Thus, training with the robust losses appears to have increased the power of identifying cancer-related genes.

\section{Extension to data with grouped covariates}\label{sec:aft_sgl}
% into the LASSO penalty
In some biological analysis, genes can be grouped by biological pathways. Thus, it is desirable to identify not only individual genes but gene groups that are associated with the outcome. We extend our approach by incorporating the group structure of covariates. We use the sparse group LASSO (SGL) penalty \cite{SGL} to replace the LASSO penalty in our proposed EM approach. To be more specific, suppose that $p$ covariates can be divided into $G$ groups and the $g$th group contains $p_g$ covariates, $g=1,\ldots,G$. Let $\beta_{gj}$ denote the regression coefficient of the $j$th covariate in the $g$th group and $x_{igj}$ denote the $j^{th}$ covariate in the $g^{th}$ group for the $i^{th}$ subject, $i=1,...,n$, $g=1,...,G$, $j=1,...,p_g$.
Then the objective function to be minimized can be formulated as follows:
\begin{align}\label{eq:obj_sgl}
l_n(\theta;\lambda_1,\lambda_2)=&\sum_{i=1}^{n}L(T_i,\theta)+  \lambda_1\sum_{g=1}^{G}\sqrt{p_g}\sqrt{\sum_{j=1}^{p_g}\beta_{gj}^2}
+ \lambda_2\sum_{g=1}^{G}\sum_{j=1}^{p_g}|\beta_{gj}|\notag\\
=&  \sum_{i\in {\cal C}} L(T_i,\theta)+\sum_{i\in {\cal D}}L(T_i,\theta) +\lambda_1\sum_{g=1}^{G}\sqrt{p_g}\sqrt{\sum_{j=1}^{p_g}\beta_{gj}^2}
+ \lambda_2\sum_{g=1}^{G}\sum_{j=1}^{p_g}|\beta_{gj}|,
\end{align}
where $\lambda_1$ and $\lambda_2$ are the tuning parameters. 

The only difference of \eqref{eq:obj_general} with \eqref{eq:obj_sgl} lies in the replacement of the lasso penalty by the SGL penalty. Thus, we modify Step (B2) in \textit{Algorithm B} to minimize the above objection function \eqref{eq:obj_sgl}. In the $m$th iteration, $\widehat{Q}(\theta|\theta^{(m-1)})$ is given by 
\begin{align*}
\widehat{Q}(\theta|\theta^{(m-1)})=&
\sum_{i\in C}\sum_{u>i}w_{iu}^{(m-1)}L(e_u^{(m-1)}+\alpha^{(m-1)}+X_i^T\beta^{(m-1)},\theta) +\sum_{i\in D} L(Y_i,\theta)\\ + & \lambda_1\sum_{g=1}^{G}\sqrt{p_g}\sqrt{\sum_{j=1}^{p_g}\beta_{gj}^2}
+ \lambda_2\sum_{g=1}^{G}\sum_{j=1}^{p_g}|\beta_{gj}|. 
\end{align*}

We still use three robust loss functions (the absolute loss, the Huber loss and the Tukey loss) to illustrate our approach. For the absolute loss, \eqref{eq:aft_ab} becomes
\begin{align} \label{eq:ab_sgl}
\widehat{Q}(\theta|\theta^{(m-1)})=\sum_{k=1}^{K} w_{k}^{new}| Y_{k}^{new} -\alpha - X_{k}^{new^T} \beta| +\lambda_1\sum_{g=1}^{G}\sqrt{p_g}\sqrt{\sum_{j=1}^{p_g}\beta_{gj}^2}
+ \lambda_2\sum_{g=1}^{G}\sum_{j=1}^{p_g}|\beta_{gj}|.
\end{align}
In order to minimize the objective function \eqref{eq:ab_sgl}, we first re-formulate the problem as 
\begin{align*}
\min_{\alpha,\beta,\nu,\eta,\zeta} & \left(\sum_{i=1}^{K} w_{i}^{new} \nu_i + \sum_{g=1}^{G} \lambda_1 \sqrt{p_g} \eta_g + \sum_{j=1}^{p} \lambda_2 \zeta_j \right),\\  
\text{subject to} \ \  
& -\nu_i \leq Y_i^{new}-\alpha-X_i^{new}\beta \leq \nu_i \ \ \text{ for }i=1,\ldots,K, \\
& \sqrt{\sum_{j=1}^{p_g}\beta_{gj}^2} \leq \eta_g \ \ \ \ \  \text{ for }g=1,\ldots,G, \\
& -\zeta_j \leq e_{j}^T \beta \leq \zeta_j \ \ \text{ for }j=1,\ldots,p,
\end{align*}
where $e_j$ is the $p$-dimensional vector with the $j$th component being 1 and all others being 0. Then, the solution to the minimization problem can be obtained by solving the above second-order cone programming problem.

For the Huber loss, \eqref{eq:aft_huber} becomes
\begin{align}\label{eq:huber_sgl}
\widehat{Q}(\theta|\theta^{(m-1)})=\sum_{k=1}^{K} w_{k}^{new} L^{\text{Huber(h)}}\left(\frac{Y_{k}^{new} - \alpha - X_{k}^{new^T}\beta}{d}\right) +\lambda_1\sum_{g=1}^{G}\sqrt{p_g}\sqrt{\sum_{j=1}^{p_g}\left(\frac{\beta_{gj}}{d}\right)^2}
+ \lambda_2\sum_{g=1}^{G}\sum_{j=1}^{p_g}\frac{|\beta_{gj}|}{d}.
\end{align}

For the Tukey loss,  \eqref{eq:aft_tukey} becomes
\begin{align}\label{eq:tukey_sgl}
\widehat{Q}(\theta|\theta^{(m-1)})=\sum_{k=1}^{K} w_{k}^{new} L^{\text{Tukey(t)}}\left(\frac{Y_{k}^{new} - \alpha - X_{k}^{new^T}\beta}{d}\right) +\lambda_1\sum_{g=1}^{G}\sqrt{p_g}\sqrt{\sum_{j=1}^{p_g}\left(\frac{\beta_{gj}}{d}\right)^2}
+ \lambda_2\sum_{g=1}^{G}\sum_{j=1}^{p_g}\frac{|\beta_{gj}|}{d}.
\end{align}

Following the idea of \textit{Algorithm C}, we use the following iterative coordinate descent algorithm to minimize the objective functions \eqref{eq:huber_sgl} and  \eqref{eq:tukey_sgl}.\\

\textit{Algorithm D:} The iterative coordinate descent algorithm for the the penalized weighted Huber and Tukey regressions with the SGL penlty
\begin{enumerate}
	\item[(D1)] Initialize $\widetilde{\alpha}$ and $\widetilde{\beta}$.
	
	\item[(D2)] Update the regression coefficients: 
	\begin{enumerate}
	    \item Calculate $r_k  =Y_{k}^{new} - \widetilde{\alpha} -\sum_{g=1}^{G} \sum_{j=1}^{p_g} X_{kgj} \widetilde{\beta_{gj}}, \ k=1,\ldots,K$ and  $\hat{d} = \frac{\text{WMedian}\left(|r_k-\text{WMedian}(r_k,w_{k}^{new})|, w_{k}^{new}\right)}{0.6745}$.
	    
	    \item For the $g$th ($g=1,\dots,G$) group, calculate 
	    $$Z_{gj} = 
	    2\sum_{k=1}^{K} w_{k}^{new} (\sum_{j=1}^{p_g}\frac{X_{kgj}^{new}}{\hat{d}}\widetilde{\beta_{gj}})\frac{X_{kgj}^{new}}{\hat{d}} + 
	    \sum_{k=1}^{K} w_{k}^{new} \phi'(\frac{r_k}{\hat{d}})\frac{X_{kgj}^{new}}{\hat{d}}, \ j=1,\ldots,p_g.$$
	    
	    If $\sqrt{\sum_{j=1}^{p_g} \left( S\left\{ Z_{gj},\frac{\lambda_2}{\hat{d}}\right\} \right)^2} \leq \frac{\lambda_1 \sqrt{p_g}}{\hat{d}}$, update all the coefficients of the $g$th group to be 0 ($(\widetilde{\beta_{g1}},\ldots,\widetilde{\beta_{gp_g}}) = \mathbf{0}$) and go to next group. Otherwise within the group $g$, go to steps (c-e).
	    
	    \item Calculate $r_k  =Y_{k}^{new} - \widetilde{\alpha} -\sum_{g=1}^{G} \sum_{j=1}^{p_g} X_{kgj} \widetilde{\beta_{gj}}, \ k=1,\ldots,K$ and  $\hat{d} = \frac{\text{WMedian}\left(|r_k-\text{WMedian}(r_k,w_{k}^{new})|, w_{k}^{new}\right)}{0.6745}$.

		\item Update $$\widehat{\beta_{gj}} = \frac{S\left\{ 2\sum_{k=1}^{K} w_{k}^{new} \frac{X_{kgj}^{new^2}}{\hat{d}^2}\widetilde{\beta_{gj}} + 
			\sum_{k=1}^{K} w_{k}^{new} \phi'(\frac{r_k}{\hat{d}}) \frac{X_{kgj}^{new}}{\hat{d}}, \frac{\lambda}{\hat{d}}\right\}}
		{ 2\sum_{k=1}^{K} w_{k}^{new} \frac{X_{kgj}^{new^2}}{\hat{d}^2} + \frac{\lambda_1 \sqrt{p_g}}{\hat{d} \sqrt{\sum_{g=1}^{p_g}\widetilde{\beta_{gj}}^2}}}.$$
		
		\item Set $\widetilde{\beta_{gj}} = \widehat{\beta_{gj}}$.
	\end{enumerate}
	
	\item[(D3)] Update the intercept term:
	\begin{enumerate}
		\item Calculate $r_k  =Y_{k}^{new} - \widetilde{\alpha} -\sum_{g=1}^{G} \sum_{j=1}^{p_g} X_{kgj} \widetilde{\beta_{gj}}, \ k=1,\ldots,K$ and  $\hat{d} = \frac{\text{WMedian}\left(|r_k-\text{WMedian}(r_k,w_{k}^{new})|, w_{k}^{new}\right)}{0.6745}$.
		\item Update $$\widehat{\alpha} = \widetilde{\alpha} + \frac{\sum_{k=1}^{K} w_{k}^{new} \phi'(\frac{r_k}{\hat{d}})}{2\sum_{k=1}^{K} w_{k}^{new}}\hat{d}.$$
		\item Set $\widetilde{\alpha} = \widehat{\alpha}$.
	\end{enumerate}
	
	\item[(D4)] Repeat Steps (D2) and (D3) until convergence.
	
\end{enumerate}

%is generated from the multivariate normal distribution ${N}_p(\bm{0},\bm{\Sigma})$ with $\Sigma_{ij} = 1$ for $i=j$ and $\Sigma_{ij} = 0.5$ for any $i \neq j$ where $\Sigma_{ij}$ is the component in the $i$th row and $j$th column of $\bm{\Sigma}$.
We conduct simulation studies to compare the performance of the SGL penalty with the LASSO penalty under four different loss functions. We use the same model as in Section \ref{sec:sim} to generate the data. As described in Section \ref{sec:sim}, the error terms are generated from the standard normal distribution or the mixture of two normal distributions. We consider a small number of covariates $p=30$ for simplicity. The number of groups is six and each group contains five covariates. The sample sizes for a training set, a validation set and a test set are still 200, 100 and 200, respectively. $X_i$ is generated from a multivariate normal distribution with zero mean, unit variance and pairwise correlation 0.5. The intercept $\alpha$ is set to be 1. We consider two scenarios for the regression coefficients in the important group. For the first scenario, all regression coefficients are non-zero in the important group. The regression coefficient vector $\beta$ is set to be $(\underbrace{1,1.5,2,2.5,3}_5,\underbrace{0,...,0}_{5 \times 2},\underbrace{1,-1,1,-1,1}_5,\underbrace{0,...,0}_{5 \times 2})$. The first scenario is named "all-important" (AIMP in short). For the second scenario, the only difference from the first scenario is that we replace the third non-zero coefficient in the important group by zero. The regression coefficient vector $\beta$ now becomes $(\underbrace{1,1.5,0,2.5,3}_5,\underbrace{0,...,0}_{5 \times 2},\underbrace{1,-1,0,-1,1}_5,\underbrace{0,...,0}_{5 \times 2})$. The second scenario is named "not-all-important" (NAIMP in short). $\sigma$ is chosen such that the SNR equals to 5. Likewise, the censoring times are generated from a uniform distribution that yields about $30\%$ censoring rate. The criteria for performance evaluation are the same as those used in Section \ref{sec:sim}. We use the second tuning procedure as discussed in Section \ref{sec:sim}. In order to compensate for the possible over-shrinkage caused by the double-penalty, we adopt the refitting method.

%We consider two scenarios for the regression coefficients. The first scenario is named "AIMP", we set $\beta = (1,1.5,2,2.5,3,\underbrace{0,...,0}_{5 \times 2},1,-1,1,-1,1,\underbrace{0,...,0}_{5 \times 2})$. The second scenario is named "NAIMP", we set $\beta = (1,1.5,0,2,3,\underbrace{0,...,0}_{5 \times 2},1,-1,0,1,1,\underbrace{0,...,0}_{5 \times 2})$. The number of groups is six and each group has five covariates for both scenarios. For the AIMP scenario, all covariates are important in an important group. For the NAIMP scenario, four covariates are important in an important group. $\sigma$ is chosen such that the SNR equals to 5. Likewise, the censoring times are generated from a uniform distribution that yields about $30\%$ censoring rate. The criteria for performance evaluation are the same as those used in Section \ref{sec:sim}. We use the second tuning procedure as discussed in Section \ref{sec:sim}. In order to compensate for the possible over-shrinkage caused by the double-penalty, we adopt the refitting method.

The simulation results for the standard normal error under AIMP and NAIMP scenarios are shown in Tables \ref{tab:normalaimp} and \ref{tab:normalnaimp}. The simulation results for the normal mixture errors under AIMP and NAIMP scenarios are shown in Tables \ref{tab:mixaimp} and \ref{tab:mixnaimp}.

We first look at the performance of four loss functions under the SGL penalty, the observations are similar to what we observed in Section \ref{sec:sim}. The Huber loss and the Tukey loss perform similarly to the squared-error loss when the errors follow the standard normal distribution, while substantially better than the squared-error loss in the case of errors with outliers. Then we compare the performance of the SGL penalty with the LASSO penalty. For a given loss function, the SGL penalty has higher sensitivity and specificity of selection and smaller SEE than the LASSO penalty. In addition, the SGL penalty has a little smaller PE. In a word, when a group structure exists in covariates, the SGL penalty outperforms the LASSO penalty under our proposed framework.

We also conducted simulation studies with errors following the $t$-distribution with 3 degrees of freedom ($t_3$). The results are summarized in the online Supporting Information.

\section{Conclusions}\label{sec:conclude}
In this paper, we have established a general framework for simultaneous variable selection and parameter estimation in the accelerated failure time model with arbitrary loss functions. This generality is achieved through a unified EM algorithm for minimization of the regularized loss function.
In the presence of outliers or heavy-tailed errors, robust loss functions such as the absolute loss, the Huber loss, and the Tukey loss have been shown, by both simulation and real data analysis, to outperform the traditional least squares in terms of both variable selection and prediction. 
In practice, these robust approaches are thus likely more desirable than existing procedures when the patient population is highly heterogeneous, giving rise to highly dispersed and perhaps outlying responses.

We have considered the LASSO penalty and the sparse group LASSO (SGL) penalty in the paper. Simulation studies have shown that the SGL penalty outperforms the LASSO penalty under our proposed framework when a group structure exists in covariates. In addition, generalizations to other penalty functions such as the adaptive LASSO, SCAD, and elastic net should be straightforward.\cite{alasso,SCAD,enet}

%\clearpage
%\nocite{*}% Show all bib entries - both cited and uncited; comment this line to view only cited bib entries;
%\bibliographystyle{wileyNJD-AMA}
\bibliography{ref}%

\clearpage
\begin{table}[!h]
	\renewcommand\arraystretch{1}
	\caption{Simulation results for the standard normal error with SNR=5}
	\label{tab:normalsnr5}
	\centering % used for centering table
	\begin{tabular}{r r r r r r}
		
		\hline			
		&    & Squared-error & Absolute  & Huber & Tukey\\
		\hline		
oracle	& SEE & 0.370(0.183) & 0.536(0.247) & 0.385(0.191) & 0.385(0.190) \\
		& PE & 1.949(0.139) & 1.969(0.136) & 1.951(0.139) & 1.951(0.139) \\
		&         &         &         &         &    \\
		Tuning 1          &       &         &         &         &    \\ 
(non-refit)	& SEE  &1.262(0.424) & 1.951(0.629) & 1.283(0.434) & 1.310(0.456) \\
		& PE & 1.996(0.137) & 2.071(0.156) & 2.000(0.136) & 2.002(0.138) \\
		& SEN &1(0)         & 0.999(0.012) &1(0)          & 1(0) \\
		& SPE & 0.677(0.189) & 0.562(0.236) & 0.662(0.200) & 0.659(0.198) \\        
		&         &         &         &         &    \\ 
(refit)	& SEE &0.575(0.397) & 0.995(0.602) & 0.624(0.413) & 0.638(0.419) \\
		& PE & 1.983(0.149) & 2.023(0.147) & 1.986(0.147) & 1.986(0.146) \\
		& SEN &0.994(0.033) & 0.991(0.037) & 0.992(0.035)  & 0.991(0.037) \\
		& SPE &0.950(0.110) & 0.862(0.192) & 0.942(0.129) & 0.941(0.129) \\
		&         &         &         &         &    \\ 
		&         &         &         &         &    \\ 
		
		Tuning 2    &        &         &         &         &    \\ 
(non-refit)	& SEE  &1.291(0.456) & 1.918(0.629) & 1.339(0.472) & 1.336(0.467) \\
		& PE & 2.000(0.138) & 2.064(0.150) & 2.003(0.139) & 2.004(0.139) \\
		& SEN &1(0)         & 0.999(0.012) &1(0)          & 1(0) \\
		& SPE & 0.678(0.196) & 0.585(0.218) & 0.681(0.196) & 0.668(0.193) \\        
		&         &         &         &         &    \\ 
(refit)	& SEE &0.578(0.428) & 0.913(0.543) & 0.595(0.396) & 0.612(0.416) \\
		& PE & 1.981(0.147) & 2.017(0.140) & 1.983(0.146) & 1.986(0.143) \\
		& SEN &0.994(0.033) & 0.991(0.037) & 0.994(0.033)  & 0.995(0.030) \\
		& SPE &0.948(0.130) & 0.890(0.148) & 0.950(0.099) & 0.939(0.134) \\

		\hline
		
	\end{tabular}
	
\end{table}

\begin{table}[!h]
	\renewcommand\arraystretch{1}
	\caption{Simulation results for the standard normal error with SNR=1} 
	\label{tab:normalsnr1}
	\centering % used for centering table
	\begin{tabular}{r r r r r r}
		\hline			
		&    & Squared-error & Absolute  & Huber & Tukey\\
		\hline	
oracle	& SEE & 1.660(0.835) & 2.183(1.023) & 1.678(0.841) & 1.676(0.833) \\
		& PE & 4.366(0.319) & 4.400(0.328) & 4.367(0.319) &4.368(0.319) \\
		&         &         &         &         &    \\
		Tuning 1          &       &         &         &         &    \\ 
(non-refit)	& SEE  &5.404(3.901) & 6.359(2.663) & 5.299(3.863) & 5.374(3.799) \\
		& PE & 4.393(0.321) & 4.442(0.321) & 4.393(0.317) &4.399(0.319) \\
		& SEN &0.925(0.162) & 0.930(0.125) &0.934(0.160)  & 0.931(0.160) \\
		& SPE & 0.738(0.177) & 0.662(0.251) & 0.722(0.178) & 0.727(0.178) \\        
		&         &         &         &         &    \\ 
(refit)	& SEE &5.823(6.297) & 6.318(4.397) & 5.834(6.060) & 5.609(5.627) \\
		& PE & 4.570(0.350) & 4.640(0.381) & 4.590(0.340) &4.589(0.352) \\
		& SEN &0.792(0.256) & 0.810(0.217) & 0.804(0.246)  & 0.808(0.231) \\
		& SPE &0.924(0.131) & 0.846(0.196) & 0.918(0.119) & 0.914(0.145) \\
		&         &         &         &         &    \\ 
		&         &         &         &         &    \\ 
		Tuning 2    &        &         &         &         &    \\ 
(non-refit)	& SEE  &6.684(5.419) & 7.149(4.403) & 6.678(5.465) & 6.707(5.495) \\
		& PE & 4.417(0.329) & 4.440(0.324) & 4.405(0.316) &4.408(0.319) \\
		& SEN &0.874(0.213) & 0.896(0.185) &0.878(0.212)  & 0.879(0.212) \\
		& SPE & 0.798(0.176) & 0.739(0.203) & 0.794(0.175) & 0.787(0.179) \\        
		&         &         &         &         &    \\ 
(refit)	& SEE &7.137(8.308) & 7.550(7.449) & 7.441(8.502) & 7.525(8.545) \\
		& PE & 4.574(0.359) & 4.610(0.375) & 4.584(0.352) &4.572(0.369) \\
		& SEN &0.749(0.313) & 0.751(0.289) & 0.748(0.311)  & 0.732(0.308) \\
		& SPE &0.939(0.090) & 0.900(0.130) & 0.937(0.085) & 0.938(0.124) \\

		\hline
		
	\end{tabular}
	
\end{table}

\begin{table}[!h]
	\renewcommand\arraystretch{1}
	\caption{Simulation results for the normal mixture errors with SNR=5} 
	\label{tab:mixsnr5}
	\centering % used for centering table
	\begin{tabular}{r r r r r r}
		\hline			
		&    & Squared-error & Absolute  & Huber & Tukey\\
		\hline
oracle	& SEE & 0.335(0.209) & 0.034(0.018) & 0.027(0.015) & 0.022(0.013) \\
		& PE & 1.055(0.171) & 0.906(0.143) & 0.902(0.143) & 0.897(0.143) \\
		&         &         &         &         &    \\
		Tuning 1          &       &         &         &         &   \\ 
(non-refit)	& SEE    &1.233(0.611) & 0.205(0.088) & 0.131(0.062) & 0.099(0.045) \\
		& PE    & 1.297(0.210) &0.977(0.158) & 0.951(0.154) & 0.935(0.149) \\
		& SEN    &1(0)          & 1(0)          & 1(0)         & 1(0)         \\
		& SPE    &0.632(0.207) & 0.490(0.235) & 0.616(0.171) & 0.588(0.179) \\        
		&         &         &         &         &    \\ 
(refit)	& SEE &0.529(0.486) & 0.045(0.034) & 0.036(0.027) & 0.028(0.019) \\
		& PE  & 1.138(0.239) &0.914(0.141) & 0.908(0.140) & 0.901(0.141) \\
		& SEN &0.989(0.047) & 1(0)          & 1(0)          & 1(0)         \\
		& SPE & 0.967(0.073) & 0.937(0.116) & 0.960(0.098) & 0.964(0.074) \\
		&         &         &         &         &    \\ 
		&         &         &         &         &    \\ 
		Tuning 2    &        &         &         &         &    \\ 
(non-refit)	& SEE    &1.163(0.609) & 0.205(0.089) & 0.123(0.056) & 0.095(0.042) \\
		& PE    & 1.291(0.210) &0.979(0.157) & 0.949(0.154) & 0.934(0.149) \\
		& SEN    &1(0)          & 1(0)          & 1(0)         & 1(0)         \\
		& SPE    &0.591(0.198) & 0.468(0.238) & 0.556(0.190) & 0.555(0.191) \\        
		&         &         &         &         &    \\ 
(refit)	& SEE &0.539(0.461) & 0.045(0.036) & 0.038(0.028) & 0.027(0.017) \\
		& PE  & 1.136(0.225) &0.914(0.141) & 0.908(0.142) & 0.900(0.142) \\
		& SEN &0.990(0.042) & 1(0)          & 1(0)          & 1(0)         \\
		& SPE & 0.958(0.087) & 0.940(0.102) & 0.952(0.106) & 0.968(0.072) \\
		\hline

	\end{tabular}
	
\end{table}

\begin{table}[!h]
	\renewcommand\arraystretch{1}
	\caption{Simulation results for the normal mixture errors with SNR=1} 
	\label{tab:mixsnr1}
	\centering % used for centering table
	\begin{tabular}{r r r r r r}
		\hline			
		&    & Squared-error & Absolute  & Huber & Tukey\\
		\hline
oracle	& SEE & 1.332(0.916) & 0.154(0.083) & 0.124(0.077) & 0.101(0.063) \\
		& PE & 2.141(0.314) & 1.862(0.255) & 1.855(0.255) & 1.846(0.256) \\
		&         &         &         &         &    \\ 
		Tuning 1          &       &         &         &         &   \\ 
(non-refit)	& SEE &4.924(4.982) & 0.769(0.298) & 0.554(0.227) & 0.433(0.177) \\
		& PE & 2.582(0.478) & 1.983(0.279) & 1.941(0.270) & 1.914(0.272) \\
		& SEN &0.950(0.175) & 1(0)          & 1(0)         & 1(0)         \\
		& SPE & 0.691(0.201) & 0.571(0.224) & 0.662(0.178) & 0.636(0.179) \\        
		&         &         &         &         &    \\ 
(refit) & SEE &4.302(5.793) & 0.260(0.247) & 0.161(0.132) & 0.120(0.077) \\
		& PE & 2.559(0.556) & 1.890(0.260) & 1.865(0.258) & 1.852(0.254) \\
		& SEN &0.856(0.215) & 0.996(0.021) & 0.999(0.012)  & 1(0)          \\
		& SPE & 0.931(0.146) & 0.916(0.156) & 0.977(0.051) & 0.977(0.045) \\
		&         &         &         &         &    \\ 
		&         &         &         &         &    \\
		Tuning 2    &        &         &         &         &    \\ 
(non-refit)	& SEE &4.272(3.301) & 0.729(0.286) & 0.508(0.210) & 0.394(0.158) \\
		& PE & 2.539(0.433) & 1.980(0.278) & 1.939(0.268) & 1.912(0.271) \\
		& SEN &0.975(0.107) & 1(0)          & 1(0)         & 1(0)         \\
		& SPE & 0.674(0.189) & 0.520(0.235) & 0.580(0.220) & 0.565(0.201) \\        
		&         &         &         &         &    \\ 
(refit) & SEE &3.646(4.381) & 0.239(0.222) & 0.187(0.161) & 0.128(0.085) \\
		& PE & 2.507(0.464) & 1.888(0.261) & 1.876(0.261) & 1.853(0.256) \\
		& SEN &0.874(0.186) & 0.998(0.018) & 1(0)          & 1(0)          \\
		& SPE & 0.932(0.139) & 0.918(0.160) & 0.947(0.122) & 0.972(0.053) \\
		\hline
		
	\end{tabular}
	
\end{table}

\begin{table}[!h]
	\renewcommand\arraystretch{1.3}
	\caption{The means and standard errors of PE for an ovarian carcinoma study} 
	\label{tab:emp_predict_error}
	\centering % used for centering table
	\begin{tabular}{ccccc}
		\hline
	&	Squared-error & Absolute  & Huber & Tukey\\
	PE&	0.847(0.098)&0.829(0.072)&0.819(0.078)&0.824(0.077)\\
		\hline
		
	\end{tabular}
	
\end{table}

\begin{table}[!h]
	\renewcommand\arraystretch{1.3}
	\caption{The top 10 most frequently selected genes by different loss functions with their selection frequencies in parentheses} 
	\label{tab:emp_top10_genes}
	\centering % used for centering table
	\begin{tabular}{rrrr}
		\hline
		Squared-error & Absolute  & Huber & Tukey\\
		\hline
		VSIG4(0.81)   & VSIG4(0.83)   & AHDC1(0.80)   &AHDC1(0.77)\\
		PPFIBP1(0.72)  & AHDC1(0.66)  & VSIG4(0.75)  & PDE1B(0.77)\\
		AHDC1(0.58)      & AKT2(0.65)   & STK4(0.65)      & STK4(0.75)\\
		WIPF2(0.50)     & HAL(0.59)    & PDE1B(0.65)      &VSIG4(0.62)\\
		NDRG3(0.49)     & CLIP3(0.57)   & PPFIBP1(0.57)      &AKT2(0.61)\\
		STK4(0.49)       &AXL(0.50)    &NDRG3(0.51)       &CLIP3(0.51)\\
		APC(0.48)       &BRD4(0.44)       &AKT2(0.50)       &PPFIBP1(0.48)\\
		HAL(0.46)    &GFPT2(0.42)     &APC(0.46)         &SYDE1(0.47)\\
		PAF1(0.43) & PPFIBP1(0.38)        & BRD4(0.42)       &NDRG3(0.45) \\
		AXL(0.36) & GALNT10(0.37)      & CLIP3(0.41)       &APC(0.42) \\	
		\hline
		
	\end{tabular}
	
\end{table}

\begin{table}[H]
	\renewcommand\arraystretch{1}
	\caption{Simulation results for the standard normal error under AIMP scenario} 
	\label{tab:normalaimp}
	\centering % used for centering table
	\begin{tabular}{r r r r r r}
		
		\hline			
		&    & Squared-error & Absolute  & Huber & Tukey\\
		\hline		
oracle	& SEE & 2.012(0.922) & 2.846(1.135) & 2.114(0.959) & 2.144(0.987) \\
		& PE & 3.204(0.238) & 3.245(0.236) & 3.210(0.239) & 3.213(0.242) \\
		&         &         &         &         &    \\ 
		LASSO    &        &         &         &         &    \\ 
(refit)	& SEE &4.774(1.671) & 5.869(1.707) & 4.853(1.720) & 4.973(1.855) \\
		& PE  &3.453(0.422) & 3.540(0.429) & 3.475(0.449) & 3.477(0.462) \\
		& SEN &0.844(0.165) & 0.798(0.171) &0.844(0.166)  & 0.836(0.181) \\
		& SPE &0.770(0.214) & 0.726(0.247) & 0.779(0.186) & 0.785(0.186) \\        
		&         &         &         &         &    \\ 
		SGL          &       &         &         &         &    \\ 
(refit)	& SEE &3.380(1.678) & 4.705(1.922) & 3.452(1.764) & 3.429(1.679) \\
		& PE  &3.196(0.229) & 3.324(0.249) & 3.199(0.231) & 3.201(0.230) \\
		& SEN &0.903(0.164) & 0.884(0.158) & 0.901(0.160)  & 0.900(0.161) \\
		& SPE &0.866(0.213) & 0.782(0.249) & 0.882(0.205) & 0.888(0.197) \\
		\hline
		
	\end{tabular}
	
\end{table}

\begin{table}[H]
	\renewcommand\arraystretch{1}
	\caption{Simulation results for the standard normal error under NAIMP scenario} 
	\label{tab:normalnaimp}
	\centering % used for centering table
	\begin{tabular}{r r r r r r}
		
		\hline			
		&    & Squared-error & Absolute  & Huber & Tukey\\
		\hline		
oracle	& SEE & 1.138(0.542) & 1.653(0.816) & 1.197(0.598) & 1.205(0.605) \\
		& PE & 2.743(0.180) & 2.776(0.183) & 2.746(0.179) & 2.747(0.178) \\
		&         &         &         &         &    \\ 
		LASSO    &        &         &         &         &    \\ 
(refit)	& SEE &2.928(1.304) & 3.668(1.305) & 2.994(1.293) & 2.913(1.237) \\
		& PE  &3.047(0.457) & 3.014(0.323) & 2.993(0.413) & 2.983(0.376) \\
		& SEN &0.866(0.158) & 0.843(0.146) &0.878(0.149)  & 0.883(0.133) \\
		& SPE &0.825(0.173) & 0.792(0.173) & 0.820(0.161) & 0.816(0.187) \\        
		&         &         &         &         &    \\ 
		SGL          &       &         &         &         &    \\ 
(refit)	& SEE &2.079(1.169) & 2.949(1.174) & 2.160(1.153) & 2.141(1.184) \\
		& PE  &2.748(0.187) & 2.831(0.188) & 2.752(0.184) & 2.750(0.183) \\
		& SEN &0.939(0.123) & 0.921(0.121) & 0.943(0.115)  & 0.941(0.117) \\
		& SPE &0.837(0.162) & 0.795(0.139) & 0.830(0.163) & 0.836(0.161) \\
		\hline
		
	\end{tabular}
	
\end{table}

\begin{table}[H]
	\renewcommand\arraystretch{1}
	\caption{Simulation results for the normal mixture errors under AIMP scenario} 
	\label{tab:mixaimp}
	\centering % used for centering table
	\begin{tabular}{r r r r r r}
		
		\hline			
		&    & Squared-error & Absolute  & Huber & Tukey\\
		\hline		
oracle	& SEE & 1.661(1.073) & 0.179(0.075) & 0.150(0.067) & 0.119(0.054) \\
		& PE & 1.752(0.309) & 1.464(0.253) & 1.457(0.249) & 1.448(0.247) \\
		&         &         &         &         &    \\ 
		LASSO    &        &         &         &         &    \\ 
(refit)	& SEE  &3.864(2.045) & 0.287(0.123) & 0.240(0.114) & 0.187(0.089) \\
		& PE & 2.271(0.732) & 1.652(0.275) & 1.633(0.279) & 1.618(0.294) \\
		& SEN &0.887(0.149) & 1(0) & 1(0)  & 1(0) \\
		& SPE & 0.740(0.223) & 0.771(0.169) & 0.820(0.175) & 0.824(0.199) \\        
		&         &         &         &         &    \\ 
		SGL          &       &         &         &         &    \\ 
(refit)	& SEE &2.841(1.873) & 0.251(0.117) & 0.204(0.100) & 0.157(0.078) \\
		& PE & 1.876(0.330) & 1.480(0.251) & 1.491(0.253) & 1.475(0.249) \\
		& SEN &0.926(0.134) & 1(0) & 1(0)  & 1(0) \\
		& SPE &0.850(0.212) & 0.790(0.219) & 0.864(0.179) & 0.870(0.177) \\
		\hline
		
	\end{tabular}
	
\end{table}

\begin{table}[H]
	\renewcommand\arraystretch{1}
	\caption{Simulation results for the normal mixture errors under NAIMP scenario} 
	\label{tab:mixnaimp}
	\centering % used for centering table
	\begin{tabular}{r r r r r r}
		
		\hline			
		&    & Squared-error & Absolute  & Huber & Tukey\\
		\hline		
oracle	& SEE & 0.956(0.635) & 0.106(0.050) & 0.079(0.039) & 0.062(0.032) \\
		& PE & 1.461(0.246) & 1.264(0.227) & 1.256(0.223) & 1.251(0.222) \\
		&         &         &         &         &    \\ 
		LASSO    &        &         &         &         &    \\ 
(refit)	& SEE  &2.473(1.569) & 0.197(0.151) & 0.165(0.096) & 0.112(0.069) \\
		& PE & 1.989(0.669) & 1.422(0.274) & 1.392(0.265) & 1.385(0.259) \\
		& SEN &0.925(0.119) & 0.999(0.012) & 1(0)  & 1(0) \\
		& SPE & 0.796(0.186) & 0.780(0.158) & 0.803(0.180) & 0.844(0.169) \\        
		&         &         &         &         &    \\ 
		SGL          &       &         &         &         &    \\ 
(refit)	& SEE &1.809(1.358) & 0.154(0.075) & 0.131(0.069) & 0.092(0.048) \\
		& PE & 1.613(0.280) & 1.276(0.226) & 1.284(0.230) & 1.273(0.226) \\
		& SEN &0.969(0.072) & 1(0) & 1(0)  & 1(0) \\
		& SPE &0.808(0.187) & 0.799(0.154) & 0.826(0.156) & 0.862(0.119) \\
		\hline
		
	\end{tabular}
	
\end{table}

\clearpage
\section*{Supporting Information}
 Additional simulation studies and results can be found in the Supporting Information section at the end of the article.

\end{document}